\documentclass[a4paper,11pt]{article}
\usepackage{jheppub} 

\title{\boldmath A New Real Tachyon Vacuum Solution in Cubic Superstring
Field Theory}

\author{E. Aldo Arroyo}
\affiliation{Centro de Ci\^{e}ncias Naturais e Humanas, Universidade Federal do ABC,\\
 Santo Andr\'{e}, 09210-170 S\~{a}o Paulo, SP, Brazil.}

\emailAdd{aldo.arroyo@ufabc.edu.br}

\abstract{We study a real tachyon vacuum solution in cubic superstring field
theory that avoids square roots and phantom terms. Using this new
solution, we evaluate the vacuum energy and obtain a result
consistent with Sen's conjecture. Additionally, we demonstrate
that the equation of motion, when contracted with the solution
itself, is satisfied.
}

\begin{document}
\maketitle
\flushbottom

\section{Introduction}

The first analytic solution to the string field equation of motion
in open bosonic string field theory \cite{Witten:1985cc} was
obtained by Schnabl \cite{Schnabl:2005gv}. His work provided a
compelling case for identifying the solution with the so-called
tachyon vacuum, an interpretation that was further supported by
subsequent developments
\cite{Ellwood:2006ba,Okawa:2006vm,Fuchs:2006hw,Ellwood:2008jh,Kawano:2008ry,Kawano:2008jv,Kiermaier:2008qu,AldoArroyo:2019hvj}.

According to Sen's conjecture \cite{Sen:1999mh,Sen:1999xm}, the
open string field equation of motion should admit a
Poincaré-invariant solution, which is identified as the tachyon
vacuum with no D-branes. This statement implies that the energy
$E$ of the true vacuum, obtained by solving the equation of
motion, must be equal in magnitude but opposite in sign to the
D-brane tension. In appropriate units, this energy corresponds to
\begin{align}
\label{introenergy1} E=-\frac{1}{2\pi^2}.
\end{align}

It is known that Schnabl's original tachyon vacuum solution
$\Psi_{Sch}$ is real in the sense that $\Psi^{\ddagger}_{Sch} =
\Psi_{Sch}$, where the double dagger denotes a composition of
Hermitian and BPZ conjugation \cite{Gaberdiel:1997ia}.

Although Schnabl's solution is real, it has some subtleties. The
solution contains a singular, projector-like state known as the
phantom term \cite{Erler:2012qr}. The presence of the phantom term
is not only necessary to analytically reproduce the correct value
of the energy (\ref{introenergy1}), but it is also required to
verify the validity of the equation of motion when contracted with
the solution itself \cite{Okawa:2006vm,Fuchs:2006hw}.

Solutions in which the phantom term does not appear, known as
simple solutions, have been constructed
\cite{Schnabl:2010tb,Erler:2009uj,Arroyo:2010fq,Zeze:2010sr,Arroyo:2010sy,Erler:2012qn}.
These solutions can be both real and non-real, though they fail to
satisfy the reality condition in certain cases. By performing
suitable gauge transformations, real solutions have been obtained
\cite{Erler:2009uj}. However, as noted in reference
\cite{Jokel:2017vlt}, the introduction of real solutions typically
involves somewhat cumbersome square roots, a feature already
present in \cite{Erler:2009uj}. In reference \cite{Jokel:2017vlt},
an alternative form of the solution without these square roots was
developed.

We would like to obtain a solution that is both real and simple,
namely, one without square roots or phantom terms. In the context
of open bosonic string field theory, given a solution that is not
necessarily real, one can use a similarity transformation to
convert it into a real solution. However, there exists an
alternative method to obtain real solutions from non-real ones
that does not rely on these similarity transformations
\cite{Jokel:2017vlt}.

Specifically, it has been shown that given a tachyon vacuum
solution $\psi$ together with its corresponding homotopy operator
$A$ \cite{Ellwood:2006ba,Ellwood:2001ig,Inatomi:2011xr}, the
string field defined by
\[
\Phi = \text{Re}(\psi) + \text{Im}(\psi) \, A \, \text{Im}(\psi)
\]
is a real solution.

We would like to emphasize that this prescription has been
developed within the framework of open bosonic string field
theory. In the present work, following the approach developed in
\cite{Jokel:2017vlt,Arroyo:2017mpd}, we demonstrate that the
construction of a real tachyon vacuum solution can also be
implemented in the case of modified cubic superstring field theory
\cite{Arefeva:1989cp}.

As we will show in detail, in the case of modified cubic
superstring field theory, starting with Gorbachev's non-real
simple tachyon vacuum solution \cite{Gorbachev:2010zz},
\begin{align}
\label{gSolsimple} \psi = (c + cKBc+B\gamma^2)\frac{1}{1+K} ,
\end{align}
and employing the prescription described in references
\cite{Jokel:2017vlt,Arroyo:2017mpd}, we can construct the
following real solution:
\begin{align}
\label{intromg1} \Phi = \frac{1}{4} \Bigg( \frac{1}{1+K} c + c
\frac{1}{1+K} + c \frac{B}{1+K} c + \frac{1}{1+K} c \frac{1}{1+K}
- \frac{1}{1+K} B \gamma^2 \frac{1}{1+K} \Bigg) + Q\text{-exact
terms},
\end{align}
where the $Q\text{-exact terms}$ are given by
\begin{align}
\label{intromg2} \frac{1}{2} \Bigg[ Q(Bc) \frac{1}{1+K} +
\frac{1}{1+K} Q(Bc) \Bigg] + \frac{1}{4} \frac{1}{1+K} Q(Bc)
\frac{1}{1+K}.
\end{align}

Using this new real tachyon vacuum solution, we compute its
corresponding energy and show that the obtained value is in
agreement with the value predicted by Sen's conjecture.
Additionally, we demonstrate that the equation of motion,
contracted with the solution itself, is satisfied.

This paper is organized as follows: In Section 2, we construct the
real tachyon vacuum solution. In Section 3, using the kinetic term
of the action, we compute the corresponding energy associated with
the solution. In Section 4, to verify that the equation of motion
contracted with the solution itself is satisfied, we compute the
cubic term of the action. In Section 5, we provide a summary and
discuss further directions of exploration. Some details related to
the computation of the cubic term are left for the appendix.

\section{Derivation of the new real solution}
In this section, we derive a new real tachyon vacuum solution in
the modified cubic superstring field theory \cite{Arefeva:1989cp}.
Using the algebraic relations satisfied by the elements of the
KBc$\gamma$ algebra, the real solution is constructed analogously
to the case of open bosonic string field theory
\cite{Jokel:2017vlt,Arroyo:2017mpd}.

Let us recall that, in the superstring case, in addition to the
basic string field elements $K$, $B$, and $c$, we include the
element $\gamma$
\cite{Arroyo:2010fq,Arroyo:2010sy,Gorbachev:2010zz,Erler:2007xt}.
These string fields satisfy the following algebraic relations
\begin{align}
&\{B,c\}=1\, , \;\;\;\;\;\;\; [B,K]=0 \, , \;\;\;\;\;\;\;
B^2=c^2=0
\, , \nonumber\\
\label{02eq2} \partial c = [K&,c] \, , \;\;\;\;\;\;\;
\partial \gamma  = [K,\gamma] \, , \;\;\;\;\;\;\; [c,\gamma]=0 \, ,
\;\;\;\;\;\;\; [B,\gamma]=0 \, ,
\end{align}
and have the BRST variations:
\begin{eqnarray}
\label{02eq3} QK=0 \, , \;\;\;\;\;\; QB=K \, , \;\;\;\;\;\;
Qc=cKc-\gamma^2 \, , \;\;\;\;\;\; Q\gamma=c \partial \gamma
-\frac{1}{2} \gamma
\partial c \, .
\end{eqnarray}

At this point, we are ready to construct the real solution. We
start by writing the following non-real solution
\cite{Gorbachev:2010zz}
\begin{eqnarray}
\label{noreal1} \psi = c \frac{1}{1+K} + \Big(B \gamma^2 + c K B c
\Big) \frac{1}{1+K},
\end{eqnarray}
which has associated an homotopy operator, $A^2=0$, given by
\begin{eqnarray}
\label{homotop1} A = B \frac{1}{1+K}
\end{eqnarray}
and satisfying,
\begin{eqnarray}
\label{homotop2} Q A + \psi A + A \psi = 1.
\end{eqnarray}
The Hermitian and BPZ conjugation of the string field $\psi$ is
given by
\begin{eqnarray}
\label{norealdag1} \psi^{\ddagger} = \frac{1}{1+K} c +
\frac{1}{1+K}\Big(B \gamma^2 + c K B c \Big) ,
\end{eqnarray}
where to derive this result (\ref{norealdag1}), we have used the
fact that $K^{\ddagger}=K$, $B^{\ddagger}=B$, $c^{\ddagger}=c$ and
$\gamma^{\ddagger}=\gamma$. From this result (\ref{norealdag1}),
it is clear that the solution $\psi$ is not real, namely $\psi
\neq \psi^{\ddagger}$.

However, if we define the real and imaginary parts of the solution
$\psi$ as follows
\begin{eqnarray}
\label{norealimre1} \text{Re}(\psi) =
\frac{\psi^{\ddagger}+\psi}{2}, \;\;\;\;\;\; \text{Im}(\psi) =
\frac{i\big(\psi^{\ddagger}-\psi\big)}{2},
\end{eqnarray}
it turns out that the string field given by
\begin{eqnarray}
\label{realnewsol1} \Phi =\text{Re}(\psi) + \text{Im}(\psi)\, A\,
\text{Im}(\psi) ,
\end{eqnarray}
satisfy the string field equation of motion, namely $Q \Phi + \Phi
\Phi =0$, and moreover this solution $\Phi$ is real, in the sense
that $\Phi^{\ddagger}=\Phi$.

Starting from equation (\ref{realnewsol1}) and employing equations
(\ref{noreal1}), (\ref{norealdag1}), and (\ref{norealimre1}), we
derive the following expression for the new real solution in terms
of the fundamental string fields $K$, $B$, $c$, and $\gamma$
\begin{align}
 \Phi = &\; \frac{1}{4}\Big( \frac{1}{1+K}c + c
\frac{1}{1+K} + c \frac{B}{1+K}  c + \frac{1}{1+K} c \frac{1}{1+K}
- \frac{1}{1+K} B \gamma^2 \frac{1}{1+K}\Big) \nonumber \\
\label{realnewsol2} & + \frac{1}{2} \Big[  Q(Bc)\frac{1}{1+K} +
\frac{1}{1+K} Q(Bc)\Big] +  \frac{1}{4} \frac{1}{1+K} Q(Bc)
\frac{1}{1+K}.
\end{align}

As we can observe, the solution (\ref{realnewsol2}) does not
contain the square root term $1/\sqrt{1+K}$, which usually appears
in the real version of the non-real solution (\ref{noreal1}).
Moreover, as we will show through explicit computations, in order
to obtain a vacuum energy value consistent with Sen's conjecture,
the real solution (\ref{realnewsol2}) does not require the
introduction of any phantom term, unlike the solution presented in
reference \cite{Erler:2007xt}, where phantom terms are necessary.

In the next section, to evaluate the vacuum energy using the real
solution, we will compute the kinetic term of the modified cubic
superstring field theory action.

\section{Computation of the kinetic term of the action and the vacuum energy}
Let us remember that the action of the modified cubic superstring
field theory \cite{Arefeva:1989cp} is given by
\begin{eqnarray}
\label{actionm1} S(\Psi) = -\frac{1}{2} \langle \Psi \, Q
\Psi\rangle - \frac{1}{3} \langle \Psi\,\Psi\,\Psi\rangle,
\end{eqnarray}
where $Q$ is the BRST operator. The string field $\Psi$ which has
ghost number 1 and picture number 0 belongs to the small Hilbert
space of the first-quantized matter+ghost open Neveu-Schwarz
superstring theory.

From the variation of the action, $\delta S(\Psi) = 0$, we can
derive the usual string field equation of motion: $Q \Psi + \Psi
\Psi = 0$. Note that if we consider a given solution $\Psi$ to
this equation of motion, the action (\ref{actionm1}) can be
reduced to the following computation:
\begin{eqnarray}
\label{actionm2} S(\Psi) = -\frac{1}{2} \langle \Psi \, Q
\Psi\rangle + \frac{1}{3} \langle \Psi \,Q \Psi\rangle = -
\frac{1}{6} \langle \Psi \,Q \Psi\rangle,
\end{eqnarray}
and therefore, the energy $E$ associated with this solution $\Psi$
is proportional to the kinetic term of the action, $\langle \Psi Q
\Psi \rangle$, namely
\begin{eqnarray}
\label{energym1} E(\Psi) =-S(\Psi) = \frac{1}{6} \langle \Psi \,Q
\Psi\rangle.
\end{eqnarray}

Therefore, to obtain the value of the vacuum energy associated
with the real tachyon vacuum solution (\ref{realnewsol2}), we need
to compute the kinetic term. To simplify the computation, it is
useful to express the solution (\ref{realnewsol2}) in the
following form:
\begin{align}
 \Phi = &\; \frac{1}{4}\Big( \frac{1}{1+K}c + c
\frac{1}{1+K} + c \frac{B}{1+K}  c + \frac{1}{1+K} c \frac{1}{1+K}
- \frac{1}{1+K} B \gamma^2 \frac{1}{1+K}\Big) \nonumber \\
\label{phibrst1} & + Q\Big\{\frac{1}{2} \Big( Bc\frac{1}{1+K} +
\frac{1}{1+K} Bc\Big) +  \frac{1}{4} \frac{1}{1+K} Bc
\frac{1}{1+K}\Big\}.
\end{align}
Plugging this solution (\ref{phibrst1}) into the kinetic term and
simplifying the resulting expression algebraically as much as
possible, we obtain
\begin{align}
\langle \Phi \, Q \Phi\rangle = & \frac{1}{32} \langle c K \gamma
\frac{1}{(1+K)^4} \gamma  \rangle + \frac{3}{32} \langle c K
\gamma \frac{1}{(1+K)^3} \gamma  \rangle - \frac{1}{16} \langle c
\frac{1}{(1+K)^4} \gamma^2 \rangle- \frac{5}{16} \langle c
\frac{1}{(1+K)^3} \gamma^2 \rangle \nonumber \\
&+\frac{1}{32}\langle c \frac{1}{(1+K)^4} \gamma K \gamma  \rangle
-\frac{1}{32}\langle c \frac{K}{(1+K)^4} \gamma^2  \rangle
+\frac{3}{32} \langle c \frac{1}{(1+K)^3} \gamma K \gamma \rangle
- \frac{3}{32} \langle c \frac{K}{(1+K)^3} \gamma^2 \rangle
\nonumber \\
&- \frac{3}{8} \langle c \frac{1}{(1+K)^2} \gamma^2 \rangle -
\frac{3}{16} \langle  \frac{B}{(1+K)^3} c K c \gamma^2 \rangle -
\frac{1}{32} \langle \frac{B}{(1+K)^2} c \frac{1}{(1+K)^2} c
\gamma K \gamma \rangle \nonumber \\
&-\frac{1}{16} \langle  \frac{B}{1+K} c \frac{1}{(1+K)^2} c \gamma
K \gamma \rangle + \frac{1}{32}\langle  \frac{B}{(1+K)^2} c
\frac{1}{(1+K)^2} \gamma  K c \gamma \rangle - \frac{1}{16}
\langle  \frac{B}{(1+K)^4} c K c \gamma^2 \rangle \nonumber \\
&+ \frac{1}{32}\langle  \frac{B}{(1+K)^2} c \frac{1}{1+K} \gamma K
c \gamma \rangle + \frac{1}{16}\langle \frac{B}{1+K} c
\frac{1}{(1+K)^2} \gamma  K c \gamma \rangle - \frac{1}{16}
\langle  \frac{B}{(1+K)^2} c \frac{1}{1+K} c \gamma^2 \rangle
\nonumber \\
 &-\frac{1}{32}\langle  \frac{B}{(1+K)^2} c
\frac{1}{1+K} c \gamma K \gamma \rangle - \frac{1}{16} \langle
\frac{BK}{(1+K)^2} c \frac{1}{(1+K)^2} c \gamma^2 \rangle+
\frac{1}{32}\langle  \frac{BK}{(1+K)^2} c \frac{1}{1+K} c \gamma^2
\rangle \nonumber \\
\label{kineticm1} &-\frac{3}{32} \langle  \frac{BK}{1+K} c
\frac{1}{(1+K)^2} c \gamma^2 \rangle.
\end{align}
All the above correlators can be computed using the following
basic correlators \cite{Arroyo:2016ajg}:
\begin{align}
\label{correm1} \langle e^{-t_1 K} c e^{-t_2 K} \gamma e^{-t_3 K}
\gamma e^{-t_4 K} \rangle & = \frac{(t_1+t_2+t_3+t_4)^2}{2 \pi^2}
\cos \Big( \frac{\pi
t_3}{t_1+t_2+t_3+t_4}\Big), \\
\label{correm2} \langle B e^{-t_1 K} c e^{-t_2 K} \gamma e^{-t_3
K} c e^{-t_4 K} \gamma \rangle & =
\frac{(t_1+t_2+t_3+t_4)(t_2+t_3)}{2 \pi^2} \cos \Big( \frac{\pi
(t_3+t_4)}{t_1+t_2+t_3+t_4}\Big), \\
\label{correm22} \langle B e^{-t_1 K} \gamma e^{-t_2 K} c e^{-t_3
K} \gamma e^{-t_4 K} c \rangle & =
\frac{(t_1+t_2+t_3+t_4)(t_3+t_4)}{2 \pi^2} \cos \Big( \frac{\pi
(t_2+t_3)}{t_1+t_2+t_3+t_4}\Big), \\
\label{correm3} \langle B e^{-t_1 K} c e^{-t_2 K} c e^{-t_3 K}
\gamma e^{-t_4 K} \gamma \rangle & =\frac{(t_1+t_2+t_3+t_4)t_2}{2
\pi^2} \cos \Big( \frac{\pi t_4}{t_1+t_2+t_3+t_4}\Big).
\end{align}
For instance, as a pedagogical illustration, let us explicitly
compute the correlator $\langle \frac{B}{(1+K)^2} c
\frac{1}{(1+K)^2} c \gamma K \gamma \rangle$, namely
\begin{align}
\label{examplem1} \langle \frac{B}{(1+K)^2} c \frac{1}{(1+K)^2} c
\gamma K \gamma \rangle = -\int_{0}^{\infty} dt_1
dt_2\,e^{-t_1-t_2}t_1t_2
\partial_{t_4} \Big[\langle B e^{-t_1 K} c e^{-t_2 K} \gamma e^{-t_3
K} c e^{-t_4 K} \gamma \rangle \Big] \Big{|}_{t_3=t_4=0}.
\end{align}
Employing equation (\ref{correm2}), we obtain
\begin{align}
\label{examplem2}
\partial_{t_4} \Big[\langle B e^{-t_1 K} c e^{-t_2 K} \gamma e^{-t_3
K} c e^{-t_4 K} \gamma \rangle \Big] \Big{|}_{t_3=t_4=0} =
\frac{t_2}{2 \pi^2}.
\end{align}
Plugging this result (\ref{examplem2}) into equation
(\ref{examplem1}), we get
\begin{align}
\label{examplem3} \langle \frac{B}{(1+K)^2} c \frac{1}{(1+K)^2} c
\gamma K \gamma \rangle = -\frac{1}{2 \pi^2}\int_{0}^{\infty}
dt_1\int_{0}^{\infty} dt_2\,e^{-t_1-t_2}t_1t_2^2 = -
\frac{1}{\pi^2} .
\end{align}
Performing similar computations for the remaining terms on the
right-hand side of equation (\ref{kineticm1}) and summing the
results, the value of the kinetic term turns out to be
\begin{align}
\label{kineticresult1} \langle \Phi \, Q \Phi\rangle = -
\frac{3}{\pi^2}.
\end{align}
And hence, the energy (\ref{energym1}) associated with the new
real tachyon vacuum solution $\Phi$ has the value
\begin{eqnarray}
\label{energyvacm1} E(\Phi)  = \frac{1}{6} \langle \Phi \,Q
\Phi\rangle = \frac{1}{6}\Big(- \frac{3}{\pi^2}\Big)= -
\frac{1}{2\pi^2},
\end{eqnarray}
which is in agreement with Sen's conjecture (\ref{introenergy1}).

Note that, in order to obtain the result (\ref{energyvacm1}), we
have assumed that the equation of motion, when contracted with the
solution itself, is satisfied, namely
\begin{equation}
\langle \Phi\,(Q\Phi+\Phi \Phi) \rangle=0.
\end{equation}
It is known that, in string field theory, there exist anomalous
solutions \( \Psi \) that, although they satisfy the equation of
motion
\begin{equation}
Q\Psi + \Psi \Psi = 0,
\end{equation}
do not fulfill the condition
\begin{equation}
\langle \Psi \, (Q\Psi + \Psi \Psi) \rangle = 0.
\end{equation}

Therefore, in the case of the proposed new solution \( \Phi \), to
demonstrate that the equation of motion contracted with the
solution itself is satisfied, we will need to compute the cubic
term of the action.

\section{Computation of the cubic term of the action}

We recall that, to obtain the value of the energy
(\ref{energyvacm1}), it has been assumed that the equation of
motion holds when contracted with the solution itself. We know
from experience with other solutions
\cite{Okawa:2006vm,Fuchs:2006hw,Arroyo:2010sy,Arroyo:2009ec} that
this assumption is not a trivial one. In general, a priori, there
is no justification for assuming the validity of
\begin{align}
\label{equationmotionm1} \langle \Phi \, Q\Phi \rangle + \langle
\Phi \, \Phi \, \Phi \rangle = 0
\end{align}
without an explicit calculation. Therefore, the cubic term of the
action must be evaluated.

We have already computed the kinetic term, and the result of this
computation is shown in equation (\ref{kineticresult1}). Thus, for
equation (\ref{equationmotionm1}) to be valid, we must show that
\begin{align}
\label{cubictermmm1} \langle  \Phi \, \Phi \, \Phi \rangle =
\frac{3}{\pi^2}.
\end{align}

Plugging the solution (\ref{phibrst1}) into the cubic term, after
lengthy algebraic manipulations\footnote{For this task, in
addition to employing relations (\ref{02eq2}) and (\ref{02eq3}),
we have used some identities that the correlators satisfy. For
instance, using these identities, we can show that the correlator
$\langle \frac{B}{1+K} c \frac{1}{1+K} c \frac{1}{1+K}
\gamma^2\rangle$ is equivalent to $\langle \frac{B}{(1+K)^2} c
\frac{1}{1+K} c \gamma^2 \rangle$. The complete list of these
identities can be found in Appendix A.}, we arrive at
\begin{align}
\langle \Phi \, \Phi \, \Phi \rangle =& \frac{1}{8} \langle c
\frac{1}{1+K} \gamma^2 \rangle+\frac{1}{8} \langle c
\frac{K}{(1+K)^2} \gamma^2 \rangle +\frac{5}{16} \langle c
\frac{1}{(1+K)^2} \gamma^2 \rangle +\frac{1}{4} \langle c
\frac{K}{(1+K)^3 } \gamma^2 \rangle \nonumber \\
& + \frac{1}{4} \langle c \frac{1}{(1+K)^3 } \gamma^2 \rangle +
\frac{1}{4} \langle \frac{B K^2}{1+K} c \frac{1}{(1+K)^2} c
\gamma^2 \rangle + \langle \frac{B K}{1+K} c \frac{1}{1+K} c
\gamma^2 \rangle \nonumber \\
& + \frac{5}{16}\langle \frac{B K}{1+K} c \frac{1}{(1+K)^2} c
\gamma^2 \rangle -\frac{1}{4} \langle \frac{B}{1+K} c
\frac{K}{1+K} c \gamma^2 \rangle + \frac{1}{8} \langle
\frac{B}{1+K} c \frac{1}{1+K} c \gamma^2 \rangle \nonumber \\
&-\frac{1}{4} \langle \frac{B}{1+K} c \frac{K^2}{(1+K)^2} c
\gamma^2 \rangle - \frac{13}{16} \langle \frac{B}{1+K} c
\frac{K}{(1+K)^2} c \gamma^2 \rangle - \frac{9}{16}\langle
\frac{B}{1+K} c \frac{1}{(1+K)^2} c \gamma^2 \rangle \nonumber \\
& + \frac{1}{4} \langle \frac{BK}{(1+K)^2} c \frac{K}{(1+K)^2} c
\gamma^2 \rangle +\frac{17}{16}\langle \frac{BK}{(1+K)^2} c
\frac{1}{1+K} c \gamma^2 \rangle +\langle \frac{BK}{(1+K)^2} c
\frac{K}{1+K} c \gamma^2 \rangle \nonumber \\
&+ \frac{5}{16}\langle \frac{B}{(1+K)^2} c \frac{K}{1+K} c
\gamma^2 \rangle - \frac{1}{4}\langle \frac{B}{(1+K)^2} c
\frac{K^2}{1+K} c \gamma^2 \rangle +\frac{3}{8}\langle
\frac{BK}{(1+K)^2} c \frac{1}{(1+K)^2} c \gamma^2 \rangle
\nonumber \\
&- \frac{3}{8} \langle \frac{B}{(1+K)^2} c \frac{K}{(1+K)^2} c
\gamma^2 \rangle- \frac{1}{4}\langle \frac{B}{(1+K)^2} c
\frac{K^2}{(1+K)^2} c \gamma^2 \rangle + \frac{3}{4}\langle
\frac{B}{(1+K)^2} c \frac{1}{1+K} c \gamma^2 \rangle \nonumber \\
&- \frac{1}{16} \langle \frac{BK^2}{(1+K)^3} c \frac{1}{(1+K)^2} c
\gamma^2 \rangle- \frac{1}{4}\langle \frac{BK^2}{(1+K)^3} c
\frac{1}{1+K} c \gamma^2 \rangle - \frac{1}{2}\langle
\frac{B}{(1+K)^2} c K c \gamma^2 \rangle \nonumber \\
&+ \frac{1}{8} \langle \frac{BK}{(1+K)^3} c \frac{K}{(1+K)^2} c
\gamma^2 \rangle - \frac{1}{4} \langle \frac{BK}{(1+K)^3} c
\frac{1}{1+K} c \gamma^2 \rangle + \frac{3}{8}\langle
\frac{BK}{(1+K)^3} c \frac{K}{1+K} c \gamma^2 \rangle \nonumber \\
&+ \frac{1}{4}\langle \frac{B}{(1+K)^3} c \frac{K}{1+K} c \gamma^2
\rangle - \frac{1}{8}\langle \frac{B}{(1+K)^3} c \frac{K^2}{1+K} c
\gamma^2 \rangle - \frac{3}{4}\langle \frac{BK}{(1+K)^3} c K c
\gamma^2 \rangle \nonumber \\
& - \frac{1}{4}\langle \frac{BK}{(1+K)^4} c K c \gamma^2 \rangle -
\frac{7}{8}\langle \frac{B}{(1+K)^3} c K c \gamma^2 \rangle -
\frac{1}{16}\langle \frac{B}{(1+K)^3} c \frac{K^2}{(1+K)^2} c
\gamma^2 \rangle \nonumber \\
\label{cubicmm1} &-\frac{1}{8}\langle B c \frac{K}{(1+K)^2} c
\gamma^2 \rangle - \frac{1}{8}\langle B c \frac{1}{1+K} c \gamma^2
\rangle - \frac{1}{4}\langle \frac{B}{(1+K)^4} c K c \gamma^2
\rangle - \frac{3}{16} \langle B c \frac{1}{(1+K)^2} c \gamma^2
\rangle.
\end{align}
All the correlators appearing in the evaluation of the cubic term
(\ref{cubicmm1}) can be computed using the basic correlators
(\ref{correm1})-(\ref{correm3}). Employing these results and
summing all terms, we obtain the value of the cubic term
\begin{align}
\label{valuecubicm1} \langle  \Phi \, \Phi \, \Phi \rangle =
\frac{3}{\pi^2}.
\end{align}
This result (\ref{valuecubicm1}) is in agreement with the expected
value (\ref{cubictermmm1}). Since we have explicitly shown that
the equation of motion is satisfied when contracted with the
solution itself, it is guaranteed that the energy associated with
the solution \( \Phi \) is directly proportional to the kinetic
term. Namely, the explicit calculation of the cubic term confirms
the result shown in equation (\ref{energyvacm1}).

\section{Summary and discussion}
We have presented a new real tachyon vacuum solution in modified
cubic superstring field theory \cite{Arefeva:1989cp}. Using this
solution, we computed its corresponding energy and found that the
result agrees with Sen's conjecture \cite{Sen:1999mh,Sen:1999xm}.
Moreover, we demonstrated that the solution satisfies the equation
of motion in a non-trivial way when contracted with itself.

In the context of both bosonic \cite{Witten:1985cc} and modified
cubic superstring field theory \cite{Arefeva:1989cp}, this study
opens the way to constructing real solutions from non-real ones.
For example, in the bosonic case, the real version of the
regularized identity-based solution \cite{Zeze:2010sr}, obtained
via the standard similarity transformation, contains square roots,
making the calculation of the energy cumbersome
\cite{AldoArroyo:2011gx}.

In the case of superstring field theory, there are also solutions
such as the identity-based solution \cite{Arroyo:2010sy} and the
so-called half-brane solution \cite{Erler:2010pr}. However, as in
the bosonic case, the real versions of these solutions contain
square roots or phantom terms. Using the prescription presented in
this paper, it should be possible to find an alternative real
version of these solutions.

Regarding Berkovits's non-polynomial open superstring field theory
\cite{Berkovits:1995ab}, since the mathematical formulation of
this theory shares the same algebraic structure as modified cubic
superstring field theory, the results presented in this paper
should be extendable to the construction of new real solutions
from non-real ones.

For example, the solution presented in reference
\cite{Erler:2013wda} is not real; it would be interesting to
construct a real version of this solution that could simplify the
calculation of the vacuum energy.

\appendix
\section{List of equivalent correlators}
In this appendix, a list of equivalent correlators is shown below,
which appear in the computation of the cubic term of the action.
\begin{align}
\label{apeneq1} \langle F_1 c F_2 \gamma F_3 \gamma F_4 \rangle &=
\langle c F_1 F_2 F_4 \gamma F_3 \gamma \rangle, \\
\label{apeneq3} \langle F_1 \gamma F_2 \gamma F_3 c F_4 \rangle &=
\langle c F_1 F_3 F_4\gamma F_2 \gamma \rangle, \\
\label{negapeneq1} \langle F_1 \gamma F_2 c  F_3 \gamma F_4
\rangle &= -
\langle c F_2 F_3 \gamma F_1 F_4 \gamma \rangle, \\
\label{apeneq4} \langle B F_1 c F_2 c F_3 \gamma  F_4 \gamma
\rangle &= \langle B F_1 F_3 c F_2 c
\gamma F_4 \gamma \rangle, \\
\label{apeneq5} \langle B F_1 \gamma F_2 \gamma F_3 c F_4 c
\rangle &= \langle c F_4 c B F_1 F_3
\gamma F_2 \gamma \rangle, \\
\label{apeneq6} \langle B F_1 c F_2 \gamma F_3 \gamma F_4 c
\rangle &= \langle c B F_1 c F_2 F_4 \gamma
F_3 \gamma \rangle, \\
\label{apeneq7} \langle B F_1 c F_2\gamma F_3 c F_4 \gamma \rangle
&= \langle B F_1 c F_2 \gamma F_3 c F_4 \gamma
\rangle, \\
\label{negapeneq2} \langle B F_1 \gamma F_2 c F_3 c F_4 \gamma
\rangle &= - \langle c  F_3 c B F_2 F_4 \gamma F_1  \gamma
\rangle, \\
\label{negapeneq3} \langle B F_1 \gamma  F_2 c F_3 \gamma F_4 c
\rangle &= - \langle B F_4 c  F_1 \gamma F_2 c F_3 \gamma \rangle
- \langle c F_2 F_3 \gamma F_1 F_4  \gamma  \rangle.
\end{align}
The string field $F_i$ denotes a general function of the basic
string field $K$. Employing the expressions for the correlators
given in equations (\ref{correm1})-(\ref{correm3}), all of the
above identities can be proved.

As a non-trivial example, let us show the equivalence
(\ref{negapeneq3}). Using the representation of the functions as:
$F_i(K)=\int_{0}^{\infty} dt_i f_i (t_i) e^{-t_i K}$, the
correlator that appears on the left-hand side of equation
(\ref{negapeneq3}) can be written as
\begin{align}
 \langle B F_1 \gamma  F_2 c F_3 \gamma F_4 c
\rangle &= \int_{0}^{\infty} dt_1 dt_2 dt_3 dt_4   \langle B
f_1(t_1)e^{-t_1 K} \gamma f_2 (t_2) e^{-t_2 K} c f_3(t_3) e^{-t_3
K} \gamma f_4(t_4) e^{-t_4 K} c \rangle \nonumber \\
&= \int_{0}^{\infty} dt_1 dt_2 dt_3 dt_4 f_1 f_2 f_3 f_4  \langle
B e^{-t_1 K} \gamma  e^{-t_2 K} c  e^{-t_3 K} \gamma  e^{-t_4 K} c
\rangle \nonumber \\
\label{apepro1} &= \int_{0}^{\infty} dt_1 dt_2 dt_3 dt_4 f_1 f_2
f_3 f_4 \frac{(t_3+t_4)(t_1+t_2+t_3+t_4)}{2 \pi^2} \cos \Big(
\frac{\pi (t_2+t_3)}{t_1+t_2+t_3+t_4}\Big).
\end{align}
Performing a similar computation as the one shown above, we can
write the other two correlators, appearing on the right-hand side
of equation (\ref{negapeneq3}), as follows:
\begin{align}
\label{apepro2} \langle B F_4 c  F_1 \gamma F_2 c F_3 \gamma
\rangle &= \int_{0}^{\infty} dt_1 dt_2 dt_3 dt_4 f_1 f_2 f_3 f_4
\frac{(t_1+t_2)(t_1+t_2+t_3+t_4)}{2 \pi^2} \cos \Big( \frac{\pi
(t_2+t_3)}{t_1+t_2+t_3+t_4}\Big). \\
\label{apepro3} \langle c F_2 F_3 \gamma F_1 F_4  \gamma  \rangle
&= \int_{0}^{\infty} dt_1 dt_2 dt_3 dt_4 f_1 f_2 f_3 f_4
\frac{(t_1+t_2+t_3+t_4)^2}{2 \pi^2} \cos \Big( \frac{\pi
(t_1+t_4)}{t_1+t_2+t_3+t_4}\Big).
\end{align}
If we write: $t_1+t_4 = (t_1+t_2+t_3+t_4) - (t_2+t_3)$, it turns
out that
\begin{align}
\label{apepro4} \cos \Big( \frac{\pi
(t_1+t_4)}{t_1+t_2+t_3+t_4}\Big) = \cos \Big( \pi - \frac{\pi
(t_2+t_3)}{t_1+t_2+t_3+t_4}\Big) = - \cos \Big( \frac{\pi
(t_2+t_3)}{t_1+t_2+t_3+t_4}\Big),
\end{align}
Plugging this result (\ref{apepro4}) into equation
(\ref{apepro3}), we obtain
\begin{align}
\label{apepro5} \langle c F_2 F_3 \gamma F_1 F_4  \gamma  \rangle
= - \int_{0}^{\infty} dt_1 dt_2 dt_3 dt_4 f_1 f_2 f_3 f_4
\frac{(t_1+t_2+t_3+t_4)^2}{2 \pi^2} \cos \Big( \frac{\pi
(t_2+t_3)}{t_1+t_2+t_3+t_4}\Big).
\end{align}
Employing the results (\ref{apepro2}) and (\ref{apepro5}), we can
compute:
\begin{align}
\label{apepro6} - \langle B F_4 c  F_1 \gamma F_2 c F_3 \gamma
\rangle - \langle c F_2 F_3 \gamma F_1 F_4  \gamma  \rangle ,
\end{align}
which gives an expression equal to the one shown on the right-hand
side of equation (\ref{apepro1}), and therefore, this result
proves the validity of the equivalence (\ref{negapeneq3}).

\acknowledgments

I would like to thank Ted Erler and Max Jokel for their useful
discussions.



\begin{thebibliography}{99}

\bibitem{Witten:1985cc}
  E.~Witten, Noncommutative Geometry and String Field Theory,
  Nucl.\ Phys.\ B {\bf 268}, 253 (1986).
  doi:10.1016/0550-3213(86)90155-0

\bibitem{Schnabl:2005gv}
  M.~Schnabl, Analytic solution for tachyon condensation in open string field
theory,
  Adv.\ Theor.\ Math.\ Phys.\  {\bf 10}, no. 4, 433 (2006)
  doi:10.4310/ATMP.2006.v10.n4.a1
  [hep-th/0511286].

\bibitem{Ellwood:2006ba}
  I.~Ellwood and M.~Schnabl, Proof of vanishing cohomology at the tachyon vacuum,
  JHEP {\bf 0702}, 096 (2007)
  doi:10.1088/1126-6708/2007/02/096
  [hep-th/0606142].

\bibitem{Okawa:2006vm}
  Y.~Okawa, Comments on Schnabl's analytic solution for tachyon condensation
in Witten's open string field theory,
  JHEP {\bf 0604}, 055 (2006)
  doi:10.1088/1126-6708/2006/04/055
  [hep-th/0603159].

\bibitem{Fuchs:2006hw}
  E.~Fuchs and M.~Kroyter, On the validity of the solution of string field theory,
  JHEP {\bf 0605}, 006 (2006)
  doi:10.1088/1126-6708/2006/05/006
  [hep-th/0603195].

\bibitem{Ellwood:2008jh}
  I.~Ellwood, \textit{The closed string tadpole in open string field theory},
  JHEP 0808, 063 (2008), [arXiv:0804.1131].

\bibitem{Kawano:2008ry}
  T. Kawano, I. Kishimoto and T. Takahashi, \textit{Gauge Invariant Overlaps for Classical Solutions in Open String
Field Theory}, Nucl. Phys. B 803, 135 (2008), [arXiv:0804.1541].

\bibitem{Kawano:2008jv}
  T. Kawano, I. Kishimoto and T. Takahashi, \textit{Schnabl's Solution and Boundary States in Open String Field
Theory}, Phys. Lett. B 669, 357 (2008), [arXiv:0804.4414].

\bibitem{Kiermaier:2008qu}
  M.~Kiermaier, Y.~Okawa and B.~Zwiebach, \textit{The boundary state from open string fields}, [arXiv:0810.1737].

\bibitem{AldoArroyo:2019hvj}
E.~Aldo Arroyo and M.~Kudrna, Numerical solution for tachyon
vacuum in the Schnabl gauge, JHEP \textbf{02}, 065 (2020)
doi:10.1007/JHEP02(2020)065 [arXiv:1908.05330 [hep-th]].

\bibitem{Sen:1999mh}
  A.~Sen, Descent relations among bosonic D-branes,
  Int.\ J.\ Mod.\ Phys.\ A {\bf 14}, 4061 (1999)
  doi:10.1142/S0217751X99001901
  [hep-th/9902105].

\bibitem{Sen:1999xm}
  A.~Sen, Universality of the tachyon potential,
  JHEP {\bf 9912}, 027 (1999)
  doi:10.1088/1126-6708/1999/12/027
  [hep-th/9911116].

\bibitem{Gaberdiel:1997ia}
  M.~R.~Gaberdiel and B.~Zwiebach, Tensor constructions of open string theories. 1: Foundations,
  Nucl.\ Phys.\ B {\bf 505}, 569 (1997)
  doi:10.1016/S0550-3213(97)00580-4
  [hep-th/9705038].

\bibitem{Erler:2012qr}
  T.~Erler and C.~Maccaferri, The Phantom Term in Open String Field Theory,
  JHEP {\bf 1206}, 084 (2012)
  doi:10.1007/JHEP06(2012)084
  [arXiv:1201.5122 [hep-th]].

\bibitem{Schnabl:2010tb}
  M.~Schnabl, Algebraic solutions in Open String Field Theory - A Lightning
Review,
  Acta Polytechnica 50, no. 3 (2010) 102
  [arXiv:1004.4858 [hep-th]].


\bibitem{Erler:2009uj}
  T.~Erler and M.~Schnabl, A Simple Analytic Solution for Tachyon Condensation,
  JHEP {\bf 0910}, 066 (2009)
  doi:10.1088/1126-6708/2009/10/066
  [arXiv:0906.0979 [hep-th]].

\bibitem{Arroyo:2010fq}
  E.~A.~Arroyo, Generating Erler-Schnabl-type Solution for Tachyon Vacuum in Cubic
Superstring Field Theory,
  J.\ Phys.\ A {\bf 43}, 445403 (2010)
  doi:10.1088/1751-8113/43/44/445403
  [arXiv:1004.3030 [hep-th]].

\bibitem{Zeze:2010sr}
  S.~Zeze, Regularization of identity based solution in string field
theory,
  JHEP {\bf 1010}, 070 (2010)
  doi:10.1007/JHEP10(2010)070
  [arXiv:1008.1104 [hep-th]].

\bibitem{Arroyo:2010sy}
  E.~A.~Arroyo, Comments on regularization of identity based solutions in string
field theory,
  JHEP {\bf 1011}, 135 (2010)
  doi:10.1007/JHEP11(2010)135
  [arXiv:1009.0198 [hep-th]].

\bibitem{Erler:2012qn}
  T.~Erler and C.~Maccaferri, Connecting Solutions in Open String Field Theory with Singular
Gauge Transformations,
  JHEP {\bf 1204}, 107 (2012)
  doi:10.1007/JHEP04(2012)107
  [arXiv:1201.5119 [hep-th]].

\bibitem{Jokel:2017vlt}
  M.~Jokel, Real Tachyon Vacuum Solution without Square Roots,
  arXiv:1704.02391 [hep-th].

\bibitem{Ellwood:2001ig}
  I.~Ellwood, B.~Feng, Y.~H.~He and N.~Moeller, The Identity string field and the tachyon vacuum,
  JHEP {\bf 0107}, 016 (2001)
  doi:10.1088/1126-6708/2001/07/016
  [hep-th/0105024].

\bibitem{Inatomi:2011xr}
  S.~Inatomi, I.~Kishimoto and T.~Takahashi, Homotopy Operators and One-Loop Vacuum Energy at the Tachyon
Vacuum,
  Prog.\ Theor.\ Phys.\  {\bf 126}, 1077 (2011)
  doi:10.1143/PTP.126.1077
  [arXiv:1106.5314 [hep-th]].

\bibitem{Arroyo:2017mpd}
E.~A.~Arroyo, Comments on real tachyon vacuum solution without
square roots, JHEP \textbf{01} (2018), 006
doi:10.1007/JHEP01(2018)006 [arXiv:1706.00336 [hep-th]].

\bibitem{Arefeva:1989cp}
  I.~Y.~Arefeva, P.~B.~Medvedev and A.~P.~Zubarev, New Representation for String Field Solves the Consistency Problem
for Open Superstring Field Theory,
  Nucl.\ Phys.\ B {\bf 341}, 464 (1990).
  doi:10.1016/0550-3213(90)90189-K

\bibitem{Gorbachev:2010zz}
  R.~V.~Gorbachev,
  \textit{New solution of the superstring equation of motion},
  Theor.\ Math.\ Phys.\  {\bf 162}, 90 (2010),
  [Teor.\ Mat.\ Fiz.\  {\bf 162}, 106 (2010)].

\bibitem{Erler:2007xt}
  T.~Erler, Tachyon Vacuum in Cubic Superstring Field Theory,
  JHEP {\bf 0801}, 013 (2008)
  doi:10.1088/1126-6708/2008/01/013
  [arXiv:0707.4591 [hep-th]].

\bibitem{Arroyo:2016ajg}
  E.~A.~Arroyo, A singular one-parameter family of solutions in cubic superstring
field theory,
  JHEP {\bf 1605}, 013 (2016)
  doi:10.1007/JHEP05(2016)013
  [arXiv:1602.00059 [hep-th]].

\bibitem{Arroyo:2009ec}
  E.~A.~Arroyo, Cubic interaction term for Schnabl's solution using Pade
approximants,
  J.\ Phys.\ A {\bf 42}, 375402 (2009)
  doi:10.1088/1751-8113/42/37/375402
  [arXiv:0905.2014 [hep-th]].

\bibitem{AldoArroyo:2011gx}
  E.~Aldo Arroyo, Level truncation analysis of regularized identity based
solutions,
  JHEP {\bf 1111}, 079 (2011)
  doi:10.1007/JHEP11(2011)079
  [arXiv:1109.5354 [hep-th]].

\bibitem{Erler:2010pr}
  T.~Erler, Exotic Universal Solutions in Cubic Superstring Field Theory,
  JHEP {\bf 1104}, 107 (2011)
  doi:10.1007/JHEP04(2011)107
  [arXiv:1009.1865 [hep-th]].

\bibitem{Berkovits:1995ab}
  N.~Berkovits, SuperPoincare invariant superstring field theory,
  Nucl.\ Phys.\ B {\bf 450}, 90 (1995)
  Erratum: [Nucl.\ Phys.\ B {\bf 459}, 439 (1996)]
  doi:10.1016/0550-3213(95)00620-6, 10.1016/0550-3213(95)00259-U
  [hep-th/9503099].

\bibitem{Erler:2013wda}
  T.~Erler, Analytic solution for tachyon condensation in Berkovits` open
superstring field theory,
  JHEP {\bf 1311}, 007 (2013)
  doi:10.1007/JHEP11(2013)007
  [arXiv:1308.4400 [hep-th]].

\end{thebibliography}


\end{document}